\documentclass[]{spie}  

\usepackage{amsmath,amsfonts,amssymb}
\usepackage{lineno} 
\usepackage{graphicx}
\usepackage[colorlinks=true, allcolors=blue]{hyperref}
\usepackage{booktabs} 
\usepackage[raggedrightboxes]{ragged2e}
\usepackage{comment}

\title{Accurately simulating gain and clock-induced charge production in the EMCCD gain register}

\author[a]{Kevin J. Ludwick}
\affil[a]{University of Alabama in Huntsville, 301 Sparkman Dr, Huntsville, AL 35899, USA}

\begin{document}
\vspace{1.5cm}
\vspace{1.5cm}
\vspace{1.5cm}
\maketitle 
\begin{abstract}
An electron-multiplying charge-coupled device (EMCCD) is capable of precise
detections in low-signal environments, able to detect a single photon through electron multiplication. It has many applications, such as faint-target astronomy, quantum optics, molecule tracing, and others, and it will be used for faint companion detection in the Roman Telescope’s coronagraph instrument. In an EMCCD, photons hit the pixels, and photo-electrons are created; these are multiplied via impact ionization as they travel through the gain register from one gain stage to the next. A high gain means a high multiplication factor, and this is achieved through a high voltage difference across a gain stage. If the gain is high enough, the chance of clock-induced charge (CIC) production in the gain register increases. The probability distribution function governing the gain
process typically used only accounts for charge multiplication if one or more electrons enter the gain register.  I discuss my implementation of the simulation of this effect and its customization in {\tt emccd\_detect}, the EMCCD detector simulator used for the Roman Telescope.  In addition, the simulator has been updated to use the exact binomial distribution for EM gain instead of the approximate Gamma distribution usually used in the literature, which is only valid for large counts.  I also examine some EMCCD data and show through maximum likelihood estimation with {\tt CIC\_gain\_register} that the data conform better to the binomial distribution versus the approximate Gamma/Erlang distribution. The use of the modified distribution would in principle improve the fidelity of Roman's testing and lead to better  EMCCD calibration and more accurate signal extraction from a frame. 
\end{abstract}

\keywords{photon counting, probability distributions, maximum likelihood estimation (MLE), clock-induced charge (CIC), charge-coupled device (CCD), electron-multiplying charge-coupled device (EMCCD)}

\section{INTRODUCTION}
\label{sec:intro}  

An electron-multiplying charge-coupled device (EMCCD) is useful for many low-signal applications, such 
as single-molecule tracing, hyperspectral imaging, quantum optics, deep-space imaging, and expolanet studies \cite{ref5,ref3, ref2, ref4, ref7, ref6}.  The photon signal from a low-signal observation 
target that has low 
contrast must compete with noise sources such as clock-induced charge (CIC), dark current, and read noise.  
An EMCCD multiplies the photo-electron counts in each pixel via a gain register, which is a series of stages that 
each pixel's charge proceeds through before read-out, and this multiplication can enable single-photon detection and reduce the effective 
read noise to sub-electron levels.  Each stage accelerates the charges through a 
voltage difference, and the charges that enter the stage gain more charges due to 
impact ionization, so more charges leave the stage than entered.  

The Python module {\tt emccd\_detect}\ is an EMCCD detector simulator I maintain which is primarily employed for simulations for the Roman Telescope's coronagraph instrument but is customizable for any EMCCD or CCD.  The latest release has undergone a major update, including more accurate simulation of customizable primary and secondary cosmic rays, smearing due to to the absence of a shutter (as will be the case for Roman), the use of the exact binomial distribution for simulation of the EM gain register, and the simulation of CIC in the gain register, or ``partial CIC" \cite{Ludwick_CIC}, which only has a measurable effect for high gain values.  In this work, I primarily discuss the improvements to the simulation of the EM gain register, and I make some comparisons to real data. I also examine some EMCCD data and show through maximum likelihood estimation that the data conform better to the binomial distribution versus the Gamma distribution, which is only accurate for large counts. The use of the modified distribution and accounting for partial CIC would improve the fidelity of simulations, EMCCD calibration, and the accuracy of signal extraction from a frame.

\section{BACKGROUND THE STATISTICS OF ELECTRON MULTIPLICATION AND SIMULATION}
\label{background}

Let $P$ be the probability that a charge turns into two via impact ionization 
in a gain stage, and assume this probability per charge is a constant 
(or assume it is the average probability of multiplication by two in a stage).  
The amplification, or gain, is typically modeled as follows\cite{Robbins}:
\begin{equation}
g = (1+P)^M,
\end{equation}
where $g$ is the gain (factor of multiplication of the charges that exit the gain register 
relative to the number that entered) and $M$ is the number of gain stages.

The probability $p^1_M$ that $x$ charges exit a gain register with $M$ stages when a 
single charge enters the gain register is due to a Galton-Watson branching process and is given in Basden (2003)\cite{Basden}, 
which cites Matsuo (1985)\cite{Matsuo}:
\begin{align}
\label{Basden1}
& p^1_M(x) = (1-P) p^1_{M-1}(x) + P \sum_{k=0}^{x} p^1_{M-1}(x-k) p^1_{M-1}(k), ~~ x,M \geq 1 \nonumber \\
& p^1_M(0) = 0, ~~~ M \geq 1 \nonumber \\
& p^1_0(x) = \delta_{1,x}, ~~ M \geq 1.
\end{align}
The probability that a charge in a stage does not multiply is $(1-P)$, so the only 
way to get $x$ outgoing charges if no multiplication occurs in stage $r$ is if $x$ charges come into the stage $r$ from stage $(r-1)$, as represented in the first term of the top line of the previous equation.  The terms in the sum represent all the ways in which an extra charge may be added (with probability $P$).  For a given charge in a gain stage, it will either multiply, with probability $P$, or not multiply, with probability $(1-P)$.  Therefore, each possible interaction scenario will consist of a product of some number of factors of $P$ and some number of factors of $(1-P)$.  Brian Sutin (2023) \cite{Sutin} helpfully recasts this in terms of a more straightforward binomial distribution and for an arbitrary number of particles $n$ entering stage $r$ of a gain register with $M$ total stages:
\begin{equation}
p^n_r(m) = \sum_{k=m/2}^x  {k \choose m-k} (1-P)^{2k-m} P^{m-k} p^n_{r-1}(k).
\label{Sutin}
\end{equation}
This form is much easier for computation.  Sutin defines the matrix $B_{mk}$ as
\begin{equation}
B_{mk} =  {k \choose m-k} (1-P)^{2k-m} P^{m-k}, ~ k \in \left[ \left\lceil \frac{m}{2} \right\rceil, m \right],
\end{equation}
where $B_{mk}$ is a square upper-triangular matrix with dimension equal to $m$, ranging from 0 to the highest desired number of counts output from the gain register.  Because of the recursion relation in Eq. \ref{Sutin}, one can see that 
\begin{equation}
p^n_M(m) = B_{mk}^M p^n_0(k),
\end{equation}
and $p^n_0(k)$ is a column vector with row $n$ = 1.  The output $p^n_M(m)$ is then column $n$ of $B_{mk}^M$, and the probability for any integer $x$ in $[0, m]$ is row $x$ of this output.
 
One can calculate the mean expected value $E[x]$ and variance $V[x]$ for the Galton-Watson branching process from its generating function to obtain, for $n$ input particles,
\begin{equation}
\label{exact_mean}
E[x] = n(1+P)^M
\end{equation}
and 
\begin{equation}
\label{exact_var}
V[x] = n(1-P)(1+P)^{M-1} ((1+P)^M -1).
\end{equation}
This probability distribution for one incoming particle can be approximated by the exponential distribution if $M$ is large and $P$ is small\cite{Basden}:
\begin{equation}
\label{exp}
p^1_M(x) \approx \frac{e^{-x/g}}{g},
\end{equation}
which has a mean expected value of $g$, which is the gain.  Note that a high value of $g=(1+P)^M$ is still attainable for small $P$ as long as $M$ is large enough.  Convolving this with itself $n-1$ times results in the approximate expression for $n$ incoming particles\cite{Ludwick_CIC}: 
\begin{equation}
\label{Pnexact}
p^n_M(x) \approx \frac{g^{-n} e^{-\frac{x}{g}} (-n+x+2) \Gamma (n+x)}{\Gamma (n) \Gamma (x+2)}, ~~ x \geq n.
\end{equation}
For large $x$, keeping only the highest order term in $x$, one can further approximate the distribution to obtain
\begin{equation}
\label{Erlang}
p^n_M(x) \approx \frac{x^{n-1}  e^{-\frac{x}{g}}}{g^n (n-1)!}, ~~ x \geq n, ~x \gg 1.
\end{equation}
This approximate expression is convenient, especially for typical EMCCD usage in which gain is large and thus $x$ is large.  However, it is not normalized as it is, although it is approximately so.  If $x \in \mathbb{R}$ and $x \in (0, \infty)$, then this is the Erlang distribution (the Gamma distribution for integer-valued $n$), and the Erlang distribution is a continuous probability distribution function which is normalized.  Its expected mean value is $ng$, which agrees exactly with Eq. \ref{exact_mean}, while the variance is $n g^2$, which Eq. \ref{exact_var} approaches for large gain.  A comparison of the variance with respect to the Gamma distribution and that with respect to the Galton-Watson branching process results in the ``extra noise factor", consistent with that from Basden (2003)\cite{Basden} and Matsuo (1985) \cite{Matsuo}.  The extra multiplicative factor ($ENF$) to account for the true variance over what is expected from the Gamma distribution is 
\begin{equation}
ENF = 1 + \frac{V[x]}{n g^2} = \frac{1}{g} + \frac{2}{g^{1/M}} - \frac{2}{g^{1+1/M}}.
\end{equation}
This goes to 1 as $g \rightarrow \infty$, which leads to $\sqrt{2}$ when Poisson noise is included. 

\begin{figure}[h!]
\begin{center}
\includegraphics[scale=0.6]{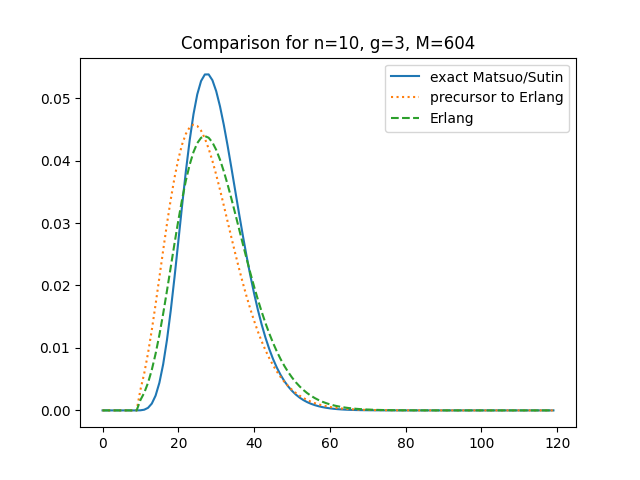}
\end{center}
\caption{ The probability functions from Eqs. \ref{Sutin} (``exact Matsuo/Sutin"), \ref{Pnexact} (``precursor to Erlang"), and \ref{Erlang} (``Erlang") for 10 incoming particles, a gain of 3, and 604 gain stages (the number of stage in Roman's EXCAM EMCCD).}
\label{Pcompare}
\end{figure}

Eqs. \ref{Sutin}, \ref{Pnexact}, and \ref{Erlang} agree reasonably well for large gains, but the peak value of the functions can differ significantly for smaller gain values.  An example comparison is shown in Fig. \ref{Pcompare}.  

\begin{figure}[h!]
\begin{center}
\includegraphics[scale=0.6]{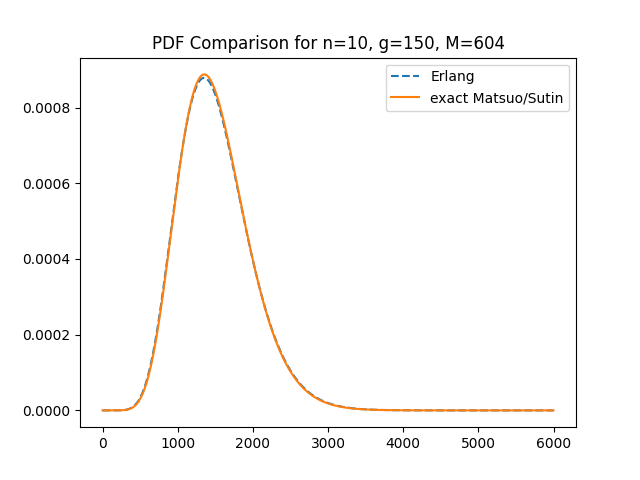}
\end{center}
\caption{ The probability functions from Eqs. \ref{Sutin} (``exact Matsuo/Sutin") and \ref{Erlang} (``Erlang") for 10 incoming particles, a gain of 150, and 604 gain stages (the number of stage in Roman's EXCAM EMCCD).}
\label{PDFn10g150M604}
\end{figure}

\begin{figure}[h!]
\begin{center}
\includegraphics[scale=0.6]{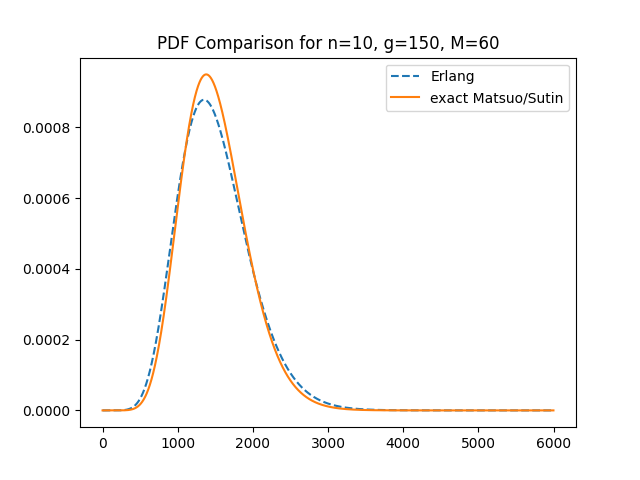}
\end{center}
\caption{ The probability functions from Eqs. \ref{Sutin} (``exact Matsuo/Sutin") and \ref{Erlang} (``Erlang") for 10 incoming particles, a gain of 150, and 60 gain stages.}
\label{PDFn10g150M60}
\end{figure}

\begin{figure}[h!]
\begin{center}
\includegraphics[scale=0.6]{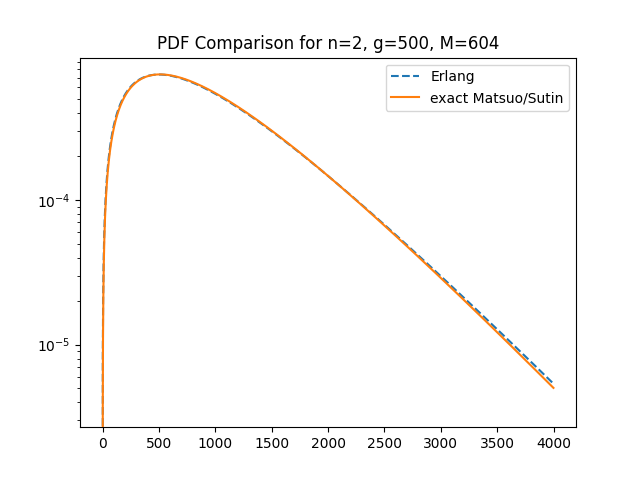}
\end{center}
\caption{ The probability functions from Eqs. \ref{Sutin} (``exact Matsuo/Sutin") and \ref{Erlang} (``Erlang") for 2 incoming particles, a gain of 500, and 604 gain stages (the number of stage in Roman's EXCAM EMCCD).}
\label{PDFLogn2g500M604}
\end{figure}

I have implemented the exact probability function from the Galton-Watson branching process in {\tt emccd\_detect}, which only used the Erlang distribution to simulate the EM gain statistics previously. This is done in Python using the {\tt binomial} random variate generator of {\tt numpy}\cite{numpy} successively $M$ times.  Figs. \ref{PDFn10g150M604}, \ref{PDFn10g150M60}, and \ref{PDFLogn2g500M604} show a few more comparisons between Eqs. \ref{Sutin} and \ref{Erlang}.  Note that the mean is the same, but the variance is visibly smaller for the exact probability function.  

I have also updated {\tt CIC\_gain\_register}\cite{Ludwick_CIC}, my Python code which can be used to perform maximum likelihood estimation (MLE) on EMCCD frames to infer the actual EM gain applied (as opposed to the commanded gain, which may differ).  The script now has the option to perform MLE using the exact probability.  Note that for large values of $ng$, the input frame data has high counts on the tail of the histogram of data, and so the number of rows and columns in the matrix $B_{nk}$ becomes very large and memory-intensive, but the execution is designed to be fairly fast.  There is a trade-off between memory intensity and computation time.  One could implement a cutoff of the higher-count data when inputting into the fitting function since the function normalizes the data accordingly for MLE.  In any case, the main purpose of the script is to perform MLE on high-gain frames potentially affected by partial CIC, and for this purpose, using the Erlang distribution as the basis for EM gain and partial CIC is sufficient.  That being said, I performed MLE on a dark frame taken at a commanded gain of 1.34, and the exact probability function was statistically preferred over the Erlang distribution, as expected.

\section{PARTIAL CIC IMPLEMENTATION}

The EM gain probability function discussed in the previous section does not account for charge multiplication as a result 
of charges that did not enter the gain register, i.e., CIC 
created in the gain register and their progeny via impact ionization.  The production 
of such charges has been studied and quantified for an EMCCD by Bush {\it et al} \cite{Bush}.  
During 
clocking in a gain stage, holes can be created on the silicon chip.  These holes 
will be accelerated due to the voltage difference in the gain stage, and they can 
collide with silicon atoms to create an electron.  This electron would then pass on 
to the remaining gain stages and multiply like all the other non-CIC electrons.  
High gain means a high voltage difference, so the effect of CIC created in the 
gain register would be most noticeable in high-gain frames of an EMCCD. 

In the MLE script {\tt CIC\_gain\_register}, the fitting functions which account for partial CIC were updated to use by default the Erlang distribution exclusively as a basis for generating the probability function inclusive of partial CIC, whereas previously they used by default the ``precursor to Erlang" as I called it in the previous section (what the code calls {\tt Pn}).  This choice was made because the exact probability function has the same mean as the Erlang distribution and is closer overall to it compared to {\tt Pn}.  One benefit of this choice is that the code was able to be sped up due to a viable additional analytic approximation, discussed below.

The probability function for partial CIC involves the convolution over all gain stages of a binomial process of spontaneous production of CIC in a given gain stage along with its progeny.  The approximate probability function for no incoming particles 
into the gain register, accounting for partial CIC, is given by\cite{Ludwick_CIC}
\begin{equation}
\label{q}
q^0_M(x) = \sum^M_{r=1} H(r,x) + (1-Q)^M \delta_{0,x},
\end{equation}
which can also be written as
\begin{equation}
q^0_M(x) =
	\begin{cases} 
		& \sum^M_{r=1} H(r,x), ~~ x>0 \\
		& (1-Q)^M, ~~ x=0.
\end{cases}
\end{equation}
$H(r,x)$ is given by 
\begin{equation}
    \label{H}
    H(r,x) \equiv \sum^{r-1}_{i=0} {r-1 \choose i} p^{i+1}_{(r+1)/2} (1-Q)^{M-(i+1)} Q^{i+1},
    \end{equation}
where $Q$ is the average probability of a clock-induced charge in a gain stage.  This function is then convolved with $p^n_M$ to obtain the probability for the EM gain register which accounts for partial CIC, and convolution with a Gaussian for read noise gives the final probability function which is used for fitting frames.  

The script previously substituted {\tt Pn} for $p^{i+1}_{(r+1)/2}$ in Eq. \ref{H}.  An approximation for $P \ll 1$ (which is accurate for a large number of gain stages) of the above expression deviated enough from the brute-force numerical sum that I decided not to use the approximation.  When substituting the Erlang distribution instead, the result is 
\begin{equation}
H(r,x) = p (1-p)^{M-1} g^{-\frac{r+1}{2 M}} e^{x \left(-g^{-\frac{r+1}{2 M}}\right)} L_{r-1}\left(\frac{g^{-\frac{r+1}{2 M}} p x}{p-1}\right),
\end{equation}
where $L_n(x)$ is the Laguerre polynomial.  I then Taylor expand this using $Q \ll 1$ and keep to third order in $Q$.  The normalization of probability functions is very important for MLE to work properly, and with {\tt Pn}, there was no viable analytic form of the normalization for Eq. \ref{q}, and it was done numerically by performing a truncated sum.  The code is faster now since there is an analytic form of the normalization constant (summing over all $x$) for Eq. \ref{q} when the 3rd-order approximation for $H(r,x)$ is used.

\begin{figure}[h!]
\begin{center}
\includegraphics[scale=0.6]{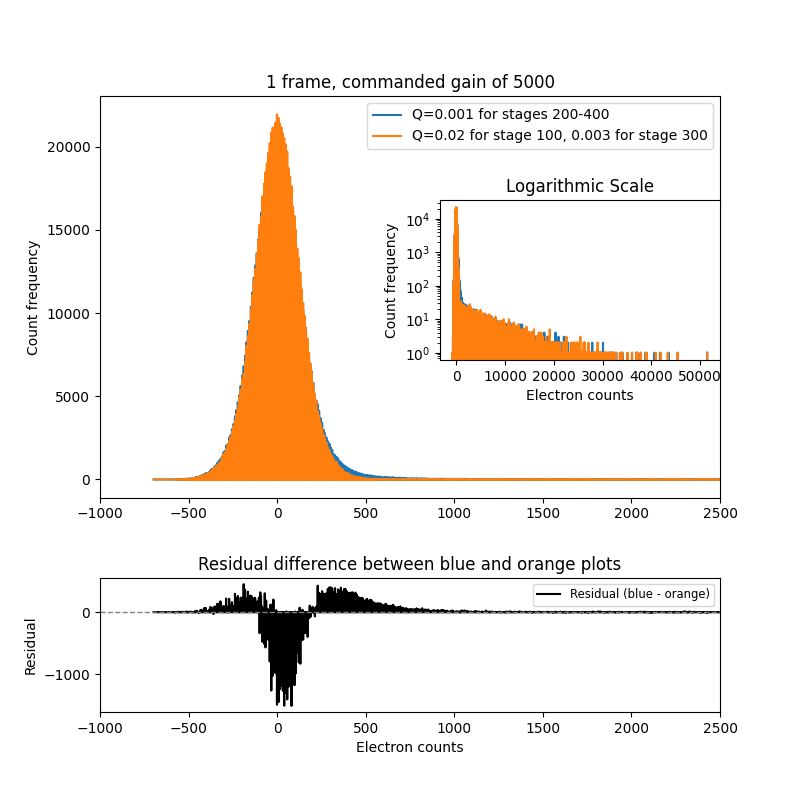}
\end{center}
\caption{ Histograms of simulated data using two different specifications of partial CIC.}
\label{pCIC_1}
\end{figure}

\begin{figure}[h!]
\begin{center}
\includegraphics[scale=0.6]{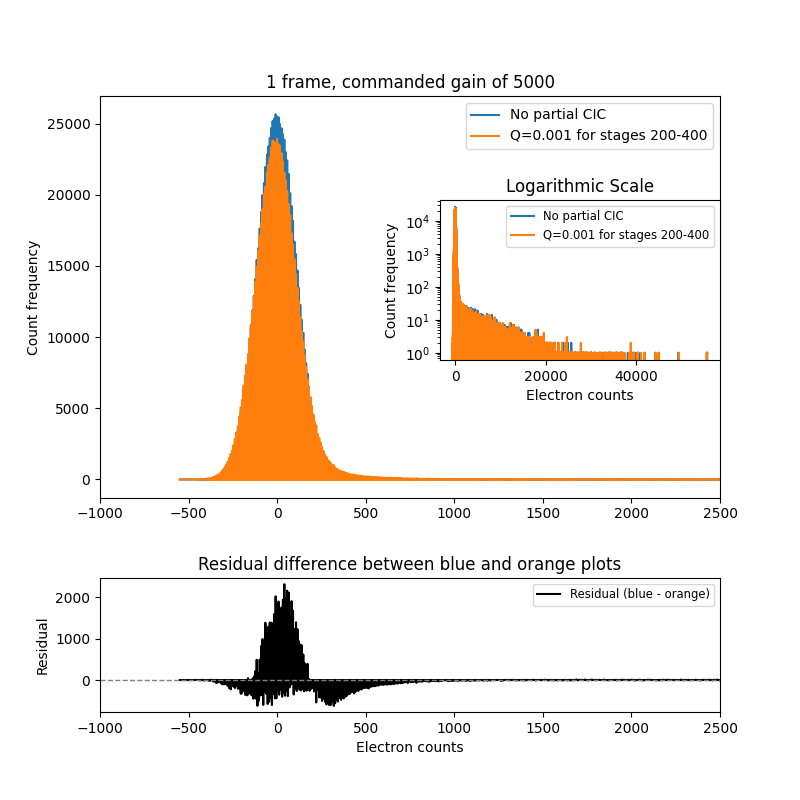}
\end{center}
\caption{Histograms of simulated data comparing no partial CIC to its inclusion.}
\label{pCIC_2}
\end{figure}

In previous work\cite{Ludwick_CIC}, I demonstrated using MLE that the inclusion of partial CIC in the probability function was statistically favored over its exclusion for real EMCCD high-gain data.  In this work, we study data simulated with the latest version of {\tt emccd\_detect} and analyze it with MLE.  Partial CIC is simulated by first generating a CIC in the gain stages according to a binomial distribution with the average chance of generation being $Q$.  Then these charges' progeny are generated using the exact gain probability function for the remaining gain stages.  The user can input either an average $Q$ value or a dictionary specifying specific $Q$ values for specific gain stages.  This is because some gain stages have been observed to be ``hot" and more highly prone to CIC production than others \cite{Bush}.  Example histograms using different partial CIC inputs are shown in Figs. \ref{pCIC_1} and \ref{pCIC_2}.  

Table \ref{result_table} shows the results of fitting a simulated dark frame using $Q = 0.001$ for gain stages 200-400 assuming various sets of parameters to fit.  The simulated image-area CIC and dark current together had a Poisson mean of $\lambda= 0.017$ electrons/pixel, and the read noise simulated was $\sigma_{rn}=110$ electrons/pixel with a mean $\mu$ of 0.  The total number of stages is 604, and the simulated EM gain was $(1+P)^M =g= 5000$.  The table shows that the fitted parameters are comparable to what was simulated.  

Simulating $Q=0.001$ for all 604 stages shifts the mean of the histogram so much that the inferred gain from simply dividing the mean value of the frame by the pre-gain electron mean of 0.017 results in an apparent gain of over 10,000.  Simulating $Q$ in stages so close to the beginning of the register is tantamount to artificially increasing the EM gain, and MLE fitting results in a higher pre-gain electron mean about 10 times higher than expected, along with a $Q$ value of about 0.01 and a gain of about 2600.  This is because of how Eq. \ref{H} was derived\cite{Ludwick_CIC}.  An approximation was made in the derivation which chose a representative, average value for the true $i$-dependent subscript of $p^{i+1}_{(r+1)/2}$ in Eq. \ref{H}.  This implies that MLE using Eq. \ref{H} is most appropriate for stages with CIC near the middle of the gain register.  This heuristic probability function is still useful for fitting real data frames, which do not have such a drastic shift in the mean due to partial CIC.  Another simulation using $Q=0.01$ only for stages 404, 400, 304, and 204 did not yield a statistical preference for the partial CIC probability function, but it did fit $Q \sim 1 \times 10^{-5}$, an effective value to accommodate this particular specification of partial CIC.  In {\tt emccd\_detect}, geared toward simulation for the Roman Telescope, the default behavior is partial CIC with $Q = P/40$ assigned to each gain stage, where $P$ is the average probability of charge multiplication in the gain stages, and this prescription aligns well with frames taken in the lab.  

\begin{table}
  \begin{center}
    \caption{The results of performing MLE on data simulated with {\tt emccd\_detect} using $Q = 0.001$ for gain stages 200-400.  All data is for 1 frame of 1024x1024 pixels, and the total number of gain stages is 604.  When read noise is not fitted for, we set $\mu=0e^-$ and $\sigma_{rn}=110e^-$.}
  \label{result_table}
     \begin{tabular}{l | p{3.9in} | p{1in}}
      \toprule 
      {Free Parameters}   & \Centering{Without Partial CIC, Full Frame} & \Centering{Log-Likelihood} \\
      \hline
      \midrule 
       \textbf{1. $\lambda$, $g$} &  \Centering{ $\hat{\lambda}=0.028016e^-, ~ \hat{g}=5000.0$ } &  \Centering{$-6.6965 \times 10^6$}\\
      \textbf{2. $\lambda$, $g$, $\mu$, $\sigma_{rn}$} &  $\hat{\lambda}=0.026116e^-, ~\hat{g}=4999.9, ~\hat{\mu}=-1.8352e^-, ~\hat{\sigma}_{rn}=118.93e^-$ &   \Centering{$-6.6906 \times 10^6$} \\
      \hline
	{Free Parameters} &  \Centering{With Partial CIC, Full Frame} &  \Centering{Log-Likelihood} \\
	\hline
	\midrule
	 \textbf{3. $\lambda$, $g$, $Q$} &   $\hat{\lambda}=0.022777e^-,~\hat{g}=5000.0, ~\hat{Q}=7.1460\times 10^{-4}$ &  \Centering{$-6.6928 \times 10^6$} \\ 
       \textbf{4. $\lambda$, $g$, $Q$, $\mu$, $\sigma_{rn}$} & $\hat{\lambda}=0.021690 e^-, ~\hat{g}=5000.0, ~\hat{Q}=8.5412 \times 10^{-4}, ~\hat{\mu}= -7.5120e^-, ~\hat{\sigma}_{rn}=117.50e^-$ &  \Centering{$-6.6866 \times 10^6$}
    \end{tabular}
  \end{center}
\end{table}

For two probability functions with the same number of fitted parameters, the one with the higher log-likelihood (closer to 0 from below) should be the preferred probability function.  When comparing the log-likelihood for two probability functions with different numbers of free parameters, I use the likelihood ratio test, where the $\chi^2$ value is 
\begin{equation}
\label{LRT}
\chi_{21}^2 = -2 (\mathcal{L}_2 - \mathcal{L}_1),
\end{equation}
where $\mathcal{L}_1$ is the log-likelihood for the probability function with $a_1$ fitted parameters and $\mathcal{L}_2$ is the log-likelihood for the probability function with $a_2<a_1$ fitted parameters.  If $\chi_{21}^2$ value is larger than the $\chi^2$ percentile corresponding to a $p$-value of $0.001$ with $a_1-a_2$ degrees of freedom, then the function with $a_1$ fitted parameters fits the data better with statistical significance.  Table \ref{chi_table} shows that there the inclusion of partial CIC in the MLE is statistically favored, as expected.  

\begin{table}
  \begin{center}
    \caption{The results of likelihood ratio tests comparing fits as numbered in Tables (\ref{result_table}).  The ratios are computed for a fit and the corresponding fit with the same fitted parameters with the addition of $Q$ for partial CIC.  We compare to the $\chi^2$ percentile value for $p=0.001$ for 1 degree of freedom, which is 10.83.  When read noise is not fitted for, we set $\mu=0e^-$ and $\sigma_{rn}=110e^-$, values from a previous calibration.}
  \label{chi_table}
     \begin{tabular}{l | p{2.4in} }
      \toprule 
      {$\chi_{ab}^2$ from likelihood ratio test}   & \Centering{Inclusion of Partial CIC Statistically Preferred} \\
      \hline
      \midrule 
       $\chi_{13}^2=7474>10.83$  &  \Centering{Yes} \\
      $\chi_{24}^2=8091>10.83$ &  \Centering{Yes} \\
    \end{tabular}
  \end{center}
\end{table}

The branching process for the exact EM gain probability function assumes that each electron can only create one additional electron per stage with probability $P$, and partial CIC is modeled so that only one charge at most per stage is created with probability $Q$.  Is it possible that the histogram of counts due to partial CIC is degenerate with that produced by a gain register in which two (or more) additional charges are produced during impact ionization with no partial CIC?  To answer this question, I used a multinomial distribution with probabilities $P_1$ for producing one additional charge per impact ionization and $P_2$ for producing two additional charges per impact ionization.  I used $P_1$ as the value corresponding to a gain of 5000 and $P_2 = P_1/10$.  This results in high counts far out on the tail of the histogram without many additional near the peak of the histogram, which is what is expected for partial CIC, and MLE fits the partial CIC probability $Q$ as 0.  I did this for several choices of $P_2$ with the same result.  Thus we can conclude that the effect seen in real data is due to partial CIC.  Physically, these two options are different; partial CIC is a phenomenon largely independent of how many particles are present in a given gain stage, whereas more than one additional charge for impact ionizations becomes more likely the more particles are present in a given gain stage.  

\section{Hot Gain Stages}
\begin{figure}[h!]
\begin{center}
\includegraphics[scale=0.6]{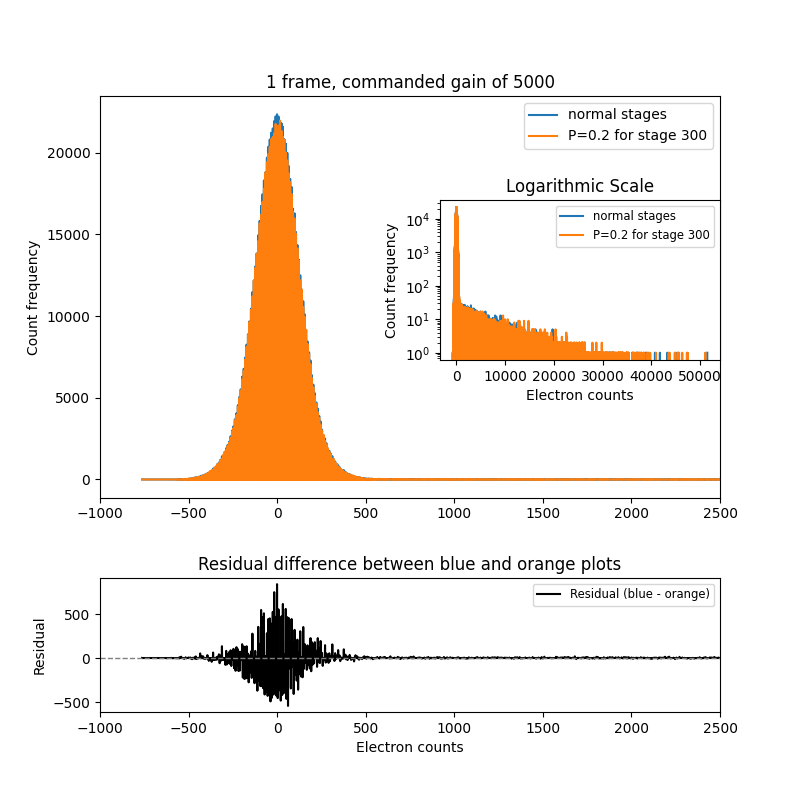}
\end{center}
\caption{Histograms of simulated data comparing no partial CIC to its inclusion.}
\label{hot_stages}
\end{figure}

EM gain is modeled as $g=(1+P)^M$, where $P$ is the average probability of multiplication via impact ionization, but a given gain stage may have a multiplication probability different from $P$ depending on the average voltage applied to that stage.  In light of this, {\tt emccd\_detect} allows one to specify the probability of multiplication on a stage by stage basis.  Fig. \ref{hot_stages} shows a comparison between a simulated frame with no deviant gain stages with one that has a hot gain stage with respect to impact ionization.  One can see that there are slightly more high-count events, and correspondingly fewer lower-count events, when there is a hot gain stage.

Physically, $Q < P$, where again $P$ is an average value, but it is possible for a ``hot" stage to have $Q$ greater than the multiplication probability for a given stage.  In fact, in {\tt emccd\_detect}, one can specify the partial CIC probability on a stage by stage basis to simulate surface charge traps in the gain register.  For example, a small $Q$ value in stage $n$ could represent charge capture, and a large $Q$ value in stage $n+2$ could represent a charge release on average two clockings later.  However, this should be coordinated with the input tail length of cosmic rays, which also simulates the effect of traps in the gain register but on a coarser scale.

\section{Conclusion}

In this paper, I discussed the implementation of more accurate EM gain simulation in {\tt emccd\_detect}, an EMCCD detector simulator applicable to any EMCCD or CCD and which is used for testing and simulation for the Roman Telescope's coronagraph instrument.  The exact binomial branching process is simulated, and that branching is also simulated for CIC produced spontaneously partway through the gain register.  The simulator can accept different $Q$ and $P$ probability specifications for each gain stage if desired.  I visually demonstrate the difference between the exact binomial probability function and the Erlang one, and I show with MLE that real data prefers the exact function over the Erlang one.  I also demonstrated a statistical preference for the probability function which accounts for partial CIC for frames simulated with partial CIC as a confirmation of the fidelity of the simulations.  The MLE script I used, {\tt CIC\_gain\_register}, was updated to be computational faster and more accurate with the Erlang distribution, which more closely approximates the exact probability function compared to the previous one used.  That script was also updated with an option to use the exact binomial probability function for MLE.  

The higher fidelity obtained in the simulations of {\tt emccd\_detect} should lead to more accurate testing for Roman.  Monte Carlo simulations could even be used for precisely characterizing the gain register of a particular EMCCD, and this would allow for high-precision threshold determination for photon counting \cite{Kevin_PC, Bijan_PC}.  The update to {\tt CIC\_gain\_register} in principle can afford better calibration of EM gain and thus better signal extraction from frames.  

\section*{Disclosures}
The authors declare that there are no financial interests, commercial affiliations, or other potential conflicts of interest that could have influenced the objectivity of this research or the writing of this paper.

This work is an expansion of these SPIE conference proceedings:  Kevin J. Ludwick, ``Accurately simulating gain and clock-induced charge production in the EMCCD gain register", Paper 14145-296,  Space Telescopes and Instrumentation 2026: Optical, Infrared, and Millimeter Wave, part of SPIE Astronomical Telescopes + Instrumentation.

\section*{CODE, DATA, AND MATERIALS AVAILABILITY}
The script that does the MLE analysis in this paper, as well as the data that was analyzed, is freely available at the public GitHub repository:  

{https://github.com/kjl0025/CIC\_gain\_register}

The module used to simulate data for this work is freely available at the public GitHub repository:

{https://github.com/roman-corgi/emccd\_detect}

\noindent All code is in Python.

\acknowledgments 

Some of this work was done under contract with the Jet Propulsion Laboratory, California Institute of Technology.
The author would like to thank Dr. Brian Sutin for alerting me to his work and for fruitful discussion.

\bibliography{report_copenhagen} 
\bibliographystyle{spiebib} 

\section*{BIOGRAPHIES}

Kevin Ludwick is a Principal Research Scientist at the University of Alabama at Huntsville, in the Center for Applied Optics.  He does research in image processing, optics calibration, and theoretical cosmology.  He earned his Ph.D. in Physics at the University of North Carolina and was a Pirrung Postdoctoral Fellow at the University of Virginia.  He was then a professor at LaGrange College.  He has served on the executive committee for the APS FECS.

\end{document}